\documentclass[pre,aps,twocolumn,showpacs,preprintnumbers,
amsmath,amssymb,superscriptaddress]{revtex4}

\usepackage{graphicx}
\usepackage{dcolumn}
\usepackage{bm}
\usepackage{hyperref} 
\usepackage{latexsym}

\begin{document}

\title{\bf Traveling length and minimal traveling time for flow \\
through percolation networks with long-range spatial correlations} 

\author{A. D. Ara\'ujo}
\affiliation{Departamento de F\'{\i}sica, Universidade Federal do
Cear\'a, 60451-970 Fortaleza, Cear\'a, Brazil.}
\affiliation{Departamento de F\'{\i}sica, Universidade Vale
do Acara\'u, 62040-370 Sobral, Cear\'a, Brazil.}
\author{A. A. Moreira}
\affiliation{Departamento de F\'{\i}sica, Universidade Federal do
Cear\'a, 60451-970 Fortaleza, Cear\'a, Brazil.}
\author{H. A. Makse}
\affiliation{Department of Physics, City College of New York, 
New York 10031-9198, USA.}
\author{H. E. Stanley}
\affiliation{Center for Polymer Studies and Department of Physics,
Boston University, Boston, Massachusetts 02215, USA.}
\author{J. S. Andrade Jr.}
\affiliation{Departamento de F\'{\i}sica, Universidade Federal do
Cear\'a, 60451-970 Fortaleza, Cear\'a, Brazil.}

\date{20 April 2002}

\begin{abstract}

We study the distributions of traveling length $l$ and minimal traveling time
$t_{min}$ through two-dimensional percolation porous media characterized by
long-range spatial correlations. We model the dynamics of fluid displacement
by the convective movement of tracer particles driven by a pressure
difference between two fixed sites (``wells'') separated by Euclidean
distance $r$. For strongly correlated pore networks at criticality, we find
that the probability distribution functions $P(l)$ and $P(t_{min})$ follow
the same scaling {\it Ansatz\/} originally proposed for the uncorrelated
case, but with quite different scaling exponents. We relate these changes in
dynamical behavior to the main morphological difference between correlated
and uncorrelated clusters, namely, the compactness of their backbones. Our
simulations reveal that the dynamical scaling exponents $d_{l}$ and $d_{t}$
for correlated geometries take values intermediate between the uncorrelated
and homogeneous limiting cases, where $l^{*}\sim r^{d_{l}}$ and
$t^{*}_{min}\sim r^{d_{t}}$, and $l^{*}$ and $t^{*}_{min}$ are the most
probable values of $l$ and $t_{min}$, respectively.

\end{abstract}

\pacs{47.55.Mh, 64.60.Ak}

\maketitle


\section{Introduction}

Fluid transport in porous media is of central importance to problems in
petroleum exploration and production \cite{Bear72,Dullien79,Sahimi95,
Ben-Avraham00}.  The geometry of an oil field can be very complex, displaying
heterogeneities over a wide range of length scales from centimeters to
kilometers \cite{King90}. The most common method of oil recovery is by
displacement. Typically water or a miscible gas (carbon dioxide or methane)
is injected in a well (or wells) to displace oil to other wells. Ultimately
the displacing fluid will break through into a production well where it must
be separated from the oil. At this point, the rate of oil production
decreases. For economic purposes, it is important to predict when the
injected fluid will break through.

When modeling the process of oil recovery, an open question is the
effect of the connectedness of the porous medium on the dynamical
process of fluid displacement. If the pore space is so poorly connected
as to be considered macroscopically heterogeneous, one expects the
overall behavior of the flowing system to display significant
anomalies. For example, it is common to investigate the physics of
disordered media at a marginal state of connectivity in terms of the
geometry of the spanning cluster at the percolation threshold
\cite{Stauffer94,Bunde96}. First, it is clearly an advantage to use the
percolation model because a comprehensive set of exactly- and
numerically-calculated critical exponents is available to describe not
only its geometrical features, but also its dynamical (transport)
properties.  Second, the application of this geometrical paradigm can be
consistently justified through ``the critical path method''
\cite{Ambegaokar71}, a powerful approximation that has been successfully
used \cite{Katz86} to estimate transport properties (e.g., permeability
and electrical conductivity) of disordered porous materials. Accordingly, the
transport in disordered media with a {\it broad} distribution of conducting
elements should be dominated by those regions where the conductances are
larger than some critical value. This value is the largest conductance such
that the set of conductances above this threshold forms a network that
preserves the global connectivity of the system. In percolation terminology,
this is equivalent to the conducting spanning cluster.

The extent to which the self-similar characteristic of the critical
percolation geometry can modify the displacement process is
unclear. Several studies have been devoted to the investigation of the
displacement process through percolation porous media at criticality
\cite{Murat86,Tian99}. More recently \cite{Lee99,Andrade00}, the
dynamics of viscous displacement through percolation clusters has been
investigated in the limiting condition of unit viscosity ratio
$m\equiv\mu _{1}/\mu _{2}$, where $\mu _{1}$ and $\mu _{2}$ are the
viscosities of the injected and displaced fluids, respectively. In this
situation, the displacement front can be modeled by tracer particles
following the streamlines of the flow, and the corresponding
distributions of shortest path and minimal traveling time closely obey a
scaling {\it Ansatz} \cite{Havlin87,Dokholyan98}. Subsequently
\cite{Andrade01}, the dynamics of viscous penetration through
two-dimensional critical percolation networks has been investigated in
the limiting case of a very large viscosity ratio,
$m\rightarrow\infty$. The results from this study indicate that the
distribution of breakthrough time follows the same scaling behavior
observed for the case $m=1$ \cite{Lee99,Andrade00}. As a consequence, 
it has been suggested \cite{Andrade01} that the process of viscous
displacement through critical percolation networks might constitute a
single universality class, independent of $m$.

The spatial distributions of porosity and permeability in real rocks are
often close to random. However, one cannot assume that the nature of their
morphological disorder is necessarily uncorrelated, i.e, the probability for
a site to be occupied is independent of the occupancy of other sites. In
fact, the permeability of some rock formations can be consistently high over
extended regions of space and low over others, characterizing in this way a
correlated structure \cite{Sahimi95}. In the case of sandstone, for example,
the permeability is not the result of an uncorrelated random process. Sand
deposition by moving water or wind (and other mechanisms of geological scale)
naturally imposes its own kind of correlations. A suitable mathematical
approach to represent the geometry of the pore spaces and the corresponding
transport properties is correlated percolation \cite{Havlin88,Havlin89,
Prakash92,Makse96a,Makse96b}. This approach has been successfully used to
model permeability fluctuations and also to explain the scale dependence of
hydrodynamic dispersion coefficients in real porous materials \cite{Makse00}.

Our aim here is to extend the investigation on the displacement dynamics
between two fluids through two-dimensional percolation clusters at
criticality \cite{Lee99,Andrade00} to the case where the pore space display
long-range spatial correlations. We focus on the case of viscous penetration
with two immiscible fluids of unit viscosity ratio ($m=1$) to study the effect
of long-range correlations on the distributions of traveling length and
minimal traveling time.

The organization of the paper is as follows. In Section II, we present the
mathematical model to simulate long-range spatial correlations and show some
geometrical features of the pore structures generated by this technique. We
also describe the dynamical model to simulate the process of viscous
displacement in porous media. We show the results in Section III and 
Section IV is discussion and summary.

\section{Model}
	
We start by describing the geometry of the disordered system studied
here. Our basic model of a porous medium is a two-dimensional site
percolation cluster at criticality \cite{Stauffer94,Bunde96} modified to
introduce correlations among the occupancy units
\cite{Havlin88,Havlin89,Prakash92,Makse96a,Makse96b}. The correlations are
induced by means of the Fourier filtering method (FFM), where a set of
random variables $u(\textbf{r})$ is introduced following a power-law
correlation function
\begin{equation}
\langle u(\textbf{r})u(\textbf{r}+\textbf{R})\rangle \propto
R^{-\gamma} \qquad\qquad [0<\gamma\leqslant 2],
\end{equation}
where $\gamma=2$ is the uncorrelated case and $\gamma\approx 0$ corresponds
to the maximum correlation. The correlated variables $u(\textbf{r})$ are used
to define the occupancy $\zeta (\textbf{r})$ of the sites
\begin{equation}
\zeta (\textbf{r})=\Theta [\phi -u(\textbf{r})]~,
\label{eq:vf1}
\end{equation}
where $\Theta$ is the Heavyside function and the parameter $\phi$ is
chosen to produce a lattice at the percolation threshold.  In Figs.~1a
and 1b we show typical backbones extracted from uncorrelated and
correlated networks.  Long-range correlations in site occupancy gives
rise to variations in the structural characteristics of the conducting
backbone \cite{Prakash92}. To illustrate this effect quantitatively,
Fig.~2 compares the fractal dimension of the conducting backbone
calculated for uncorrelated and for correlated networks with $\gamma =
0.5$. Indeed, the fractal dimension of the backbone is significantly
larger for the correlated case.

For a given correlated network at criticality, we choose two sites $A$ and
$B$ belonging to the infinite cluster and separated by a distance $r$. In oil
recovery these represent the injection and production wells (see Fig.~3). We
then extract the percolation backbone between these two points. To model
incompressible flow through this disordered system, we assume that the
lattice sites have negligible volume and the allocated bonds are homogeneous
elementary units of a porous material with constant permeability $k$ and flow 
area $a$. We also consider that the dynamics of fluid displacement is
governed by viscous forces and that $m=1$ (the invading and displaced fluids
have the same viscosity). Under these conditions and due to the strictly
convective nature of the penetration process, the velocity at each elementary
unit can be modeled in terms of Darcy's law
\begin{equation}
{v}_{ij} =\frac{k}{\mu \ell}({P}_{i}-{P}_{j})~,
\label{eq:vf2}
\end{equation}
where ${P}_{i}$ and ${P}_{j}$ are the values of pressure at sites $i$ and
$j$, respectively, and $\ell$ is the length of the bond. Conservation
at each site of the backbone leads to the following set of linear algebraic
equations:
\begin{equation}
\sum_{j} {q}_{ij}=\frac{k a}{\mu \ell}\sum_{j}({P}_{i}-{P}_{j})=0~,
\label{eq:vf3}
\end{equation}
where $q_{ij}$ is the volume flow rate through the bond and the summation is
taken over all bonds connected to a node $i$ that belongs to the cluster. 
As a macroscopic boundary condition, we impose a constant flow rate $Q$
between the injecting point $A$ and the extracting point $B$. In practice, we
apply a unit pressure drop between wells $A$ and $B$, and calculate the
solution of Eq.~(\ref{eq:vf3}) in terms of the pressure field by means of a
standard subroutine for sparse matrices. Due to the linearity of the system,
the computed velocities at each bond, $v_{ij}$, can be rescaled to give a
fixed total flow rate $Q$, independent of the distance between $A$ and $B$,
and the realization of the porous medium. This resembles more closely oil
recovery processes where constant flow is maintained instead of constant
pressure drop.

To simulate the displacement of fluid through the percolation backbone,
we first note that, under the conditions of unit viscosity ratio ($m=1$)
and for a fixed pressure drop between the wells, the pressure field remains
constant during the propagation of the invading front through the
percolation network. Another consequence of this simplifying assumption
is that the front of invading fluid in any bond $(ij)$ of the lattice
advances locally with a constant velocity equal to $v_{ij}$. This
situation can be expressed as
\begin{equation}
\frac {\partial F_{ij}}{\partial t} + v_{ij} \frac{\partial F_{ij}} 
{\partial x}=0~,
\label{eq:convective}
\end{equation}
where $F_{ij}(x,t)$ denotes the interface between invading and displaced
fluids, $t$ is time and $x$ corresponds to the local longitudinal
coordinate within each elementary unit (bond) of the porous
material. Equation~(\ref{eq:convective}) expresses the fact that the
physical system considered here is always and everywhere convective for
any value of the imposed flow rate $Q$. This behavior is entirely
analogous to the convective (non-diffusive) regime of hydrodynamical
dispersion \cite{Bear72,Dullien79}, where the unsteady transport of a
neutral tracer in a carrier fluid flowing through a porous material is
totally dominated by convective effects.  In the absence of diffusive
effects, the tracer samples the disordered medium by following the
velocity streamlines. In the general case of hydrodynamical dispersion,
however, diffusion might play a significant role. If the pore space is
sufficiently heterogeneous, local zones of small velocities can be
found, even under conditions of high overall flow rates. As a
consequence, the propagation of the tracer front in these regions may be
diffusion-like if the characteristic time for convection,
$\tau_c\equiv\ell/v$, is greater than the typical diffusion time,
$\tau_d\equiv\ell^2/D_m$, where $D_m$ is the molecular diffusion
coefficient of the tracer in the carrier fluid.

Applying the analogy between fluid displacement and the convective
propagation of a tracer through a disordered porous material, we can now
put forward a random walk picture for the front penetration of the
invading fluid. Here we follow Refs.~\cite{Lee99,Andrade00} and use the
particle-launching algorithm (PLA), where the movement of a set of
(tracer) particles is statistically dictated by the local velocity
field. In the PLA, each particle starting from the injection point $A$
can travel through the medium along a different path connected to the
recovery point $B$, taking steps of length $\ell$ and duration
$t_{ij}=\ell/v_{ij}$ (Fig.~4). The probability $p_{ij}$ that a tracer
particle at node $i$ selects an outgoing bond ${ij}$ (a bond where
$v_{ij}>0$) is proportional to the velocity of flow on that bond,
$p_{ij} \propto v_{ij}/\sum_{k}v_{ik}$, where the summation on $k$ is
over all outgoing bonds.

\section{Results}

We investigate the effect of spatial long-range correlations on the
distributions of traveling length and minimal traveling time. The {\it
traveling time} $t$ of a path $\cal {C}$ is defined as the sum of the
time steps $t_{ij}$ through each bond ${ij}$ belonging to a connected
path between $A$ and $B$
\begin{equation}
t\equiv\sum _{(ij)\in \textit {C}} t_{ij}~.
\end{equation}
The {\it traveling length} $l$ is the number of bonds present
in path $\cal C$.  Among the ensemble of all paths $\{{\cal C}\}$, we
select the path $\cal C^*$ that has the {\it minimal traveling time},
$t_{min}$,
\begin{equation}
t_{min}({\cal C^*})\equiv\min_{\bf \{ \cal C \} } {t}({\cal C})~. 
\end{equation}
This quantity corresponds to the {\it breakthrough time} of the displacing
fluid. For a given realization of the percolation network, we compute all the
traveling lengths and the minimal traveling time corresponding to the
trajectories of $10~000$ tracer particles. For a fixed value of $r$, this
operation is repeated for 10~000 network realizations of size $L \times L$,
where $L=512$, so $L\gg r$. We carried out simulations for different values
of $r$ and find that there is always a well-defined region where the
distributions of $P(l)$ and $P(t_{min})$ follow the scaling form
\cite{Havlin87}
\begin{equation}
P(z)=A_{z}\left(\frac{z}{z^{*}}\right)^{-g_{z}}f\left(\frac{z}{z^{*}}\right)~,
\label{eq:Ansatz}
\end{equation}
where $z$ denotes $l$ or $t_{min}$, $z^*$ is the maximum of the
probability distribution, the normalization constant is given by $A_{z}
\sim (z^*)^{-1}$ and the scaling function has the form
\cite{Lee99,Andrade00}
\begin{equation}
f(y)=\exp(-a_{z}y^{-\phi_{z}})~.
\label{eq1x}
\end{equation}
The exponents
$\phi_z$ and $d_z$ are related by \cite{Gennes79}
\begin{equation}
\phi_z=1/(d_z-1)~.
\label{eq2x}
\end{equation}
Note that the scaling function $f$ decreases sharply when $z$ is smaller
than $z^*$. The lower cutoff is due to the constraint, $l\ge r$.

In Figs.~4(a) and 5(a) we show log-log plots of the probability densities
$P(l)$ and $P(t_{min})$, respectively, for five different values of
``well'' separation, $r=4$, $8$, $16$, $32$, and $64$. For each curve, we
determine the characteristic size $z^*$ as the peak of the distribution
and plot $z^*$ on a double logarithmic scale. As
shown in Figs.~4(b) and 5(b), the results of our simulations indicate
that both $l^*$ and $t_{min}^*$, respectively, have  power-law
dependences on the distance $r$, $z^* \sim r^{d_z}$. The linear fit to
the data yields the exponents $d_z$ for each distribution, namely,
\begin{equation}
d_{l}=1.13 \pm 0.02 \qquad \text(correlated)
\label{eq3x}
\end{equation}
and 
\begin{equation}
d_{t}=1.75 \pm 0.03 \qquad \text(correlated). 
\label{eq4x}
\end{equation}
The same exponents reported in \cite{Lee99,Andrade00} for
the case of flow through uncorrelated percolation networks
($\gamma=2.0$) at constant flux are 
\begin{equation}
d_{l} \approx 1.21 \qquad \text(uncorrelated)
\label{eq5x}
\end{equation}
and 
\begin{equation}
d_{t} \approx 1.33 \qquad \text(uncorrelated).  
\label{eq6x}
\end{equation}
Once more, the differences in these exponents for the correlated and
uncorrelated cases can be explained in terms of the morphology of the
conducting backbone. As $\gamma$ decreases, the backbone becomes gradually
more compact \cite{Prakash92}. This distinctive feature of the correlated
geometry tends to reduce the value of $d_{l}$ and augment the value of
$d_{t}$ as the strength of the long-range correlations increases (i.e.,
$\gamma$ decreases). In the limiting case of a homogeneous system, the
corresponding exponents are $d_{l}=1$ and $d_{t}=2$ \cite{Bear72,
Andrade00,King01}.

Figures 4(c) and 5(c) show the data collapse obtained by
rescaling $l$ and $t_{min}$ by their characteristic size, $l^*$ and
$t_{min}^*$. Both scaled distributions are consistent with the scaling
form of Eq.~(\ref{eq:Ansatz}). From the least-square fit to the data in
the scaling regions, we obtain the exponents 
\begin{equation}
g_{l}=2.35 \pm 0.05 \qquad \text(correlated)
\label{eq9x}
\end{equation}
and 
\begin{equation}
g_{t}=1.89 \pm 0.04 \qquad \text(correlated). 
\label{eq10x}
\end{equation}
For uncorrelated pore networks subjected to the condition of constant
flux, the exponents are \cite{Lee99,Andrade00} 
\begin{equation}
g_{l}\approx 2.0 \qquad \text(uncorrelated)
\label{eq11x}
\end{equation}
and 
\begin{equation}
g_{t}\approx 2.0 \qquad \text(uncorrelated). 
\label{eq12x}
\end{equation}
The differences between these distribution exponents have their origins
in the different levels of compactness between correlated and
uncorrelated clusters. These results are compatible with those of
previous studies \cite{Prakash92,Makse96a}, which indicate that spatial
correlations can change other critical exponents.

\section{Discussion}

The need for a better description of the geometrical features of the
pore space has been the main conclusion of several recent experimental
and theoretical studies on transport phenomena in disordered porous
materials \cite{Sahimi95}. It is therefore necessary to examine {\it
local} aspects of the pore space morphology and relate them to the
relevant mechanisms of momentum, heat and mass transfer in order to
understand the important interplay between porous structure and
phenomenology. From a conceptual point of view, this task has been
accomplished in many works, where computational simulations
based on a detailed description of the pore space have been fairly
successful in predicting and validating known correlations among
transport properties of real porous media \cite{Kostek92,Schwartz93,
Martys94,Koponen97,Andrade97}. In the present work, we have investigated
the dynamics of immiscible fluid displacement using the framework of a
percolation model for porous media that has been specially modified to
introduce spatial long-range correlations among the occupancy units of
permeability. This model is certainly a more realistic description for
the geometry of porous rocks and should lead to a better mathematical
representation of their transport properties.

Our results on the distributions of traveling length and minimal
traveling time through correlated percolation networks show that spatial
fluctuations in rock permeability can have significant consequences on
the dynamics of fluid displacement. More precisely, we observed that the
presence of long-range correlations can substantially modify the scaling
exponents of these distributions and, therefore, their universality
class. As in previous studies on the subject \cite{Prakash92,Makse96a},
we explain this change of behavior in terms of the morphological
differences among uncorrelated and correlated pore spaces generated at
criticality. Compared to the uncorrelated structures, the backbone
clusters of the correlated cluster has a more compact geometry. The
level of compactness depends, of course, on the degree of correlations
introduced during the generation process. Moreover, our results are
consistent with the fact that the dynamical scaling exponents $d_{l}$
and $d_{t}$ obtained for correlated geometries assume values intermediate
between the uncorrelated and the homogeneous limiting cases.

This work has been supported by CNPq, FUNCAP, NSF, and BP-Amoco. We
thank S. V. Buldyrev, N. V. Dokholyan, S. Havlin, P. R. King, Y. Lee,
E. Lopez and G. Paul for helpful interactions.

\newpage

\begin{figure}
\includegraphics[width=8.0cm]{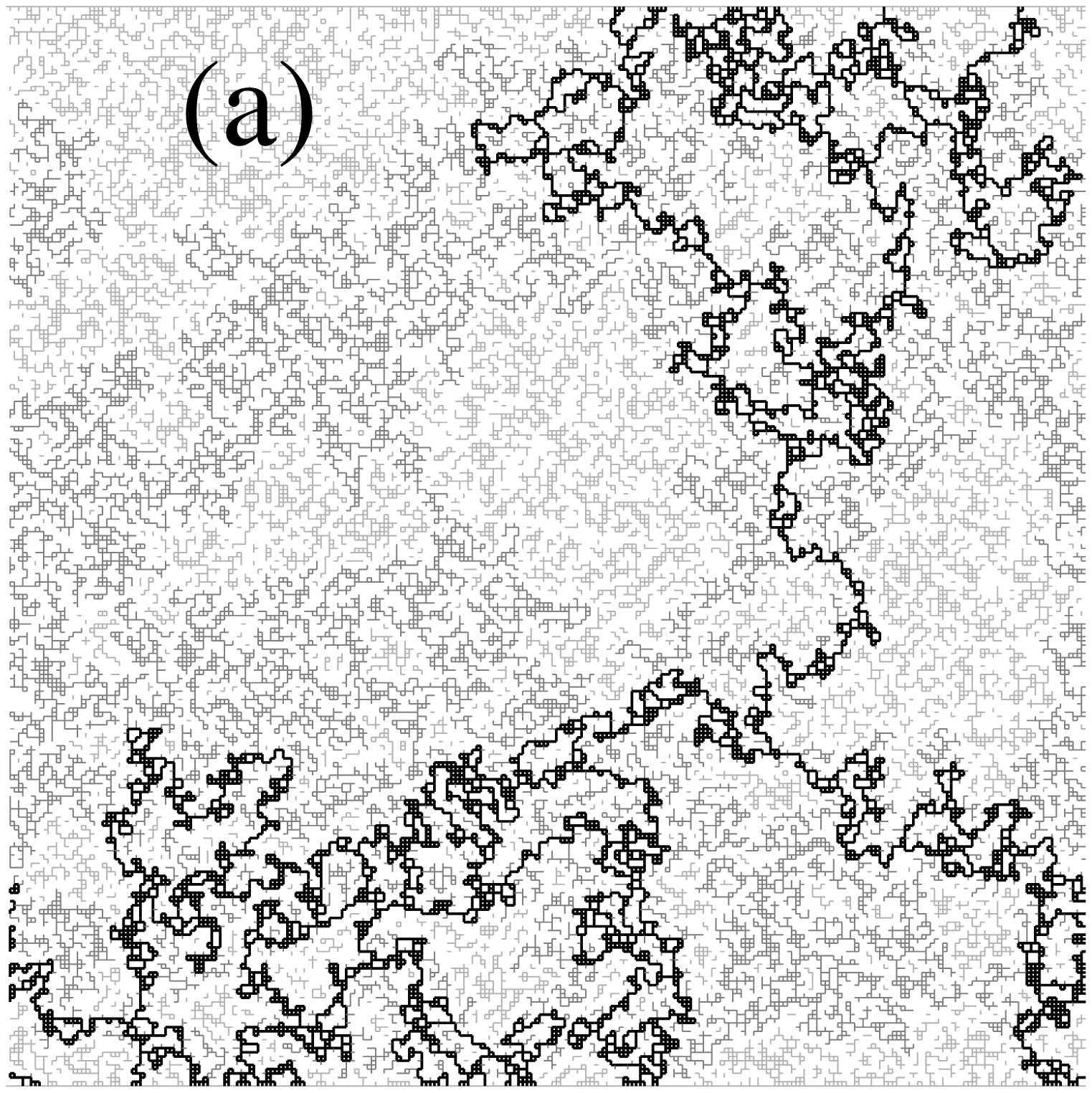}
\includegraphics[width=8.0cm]{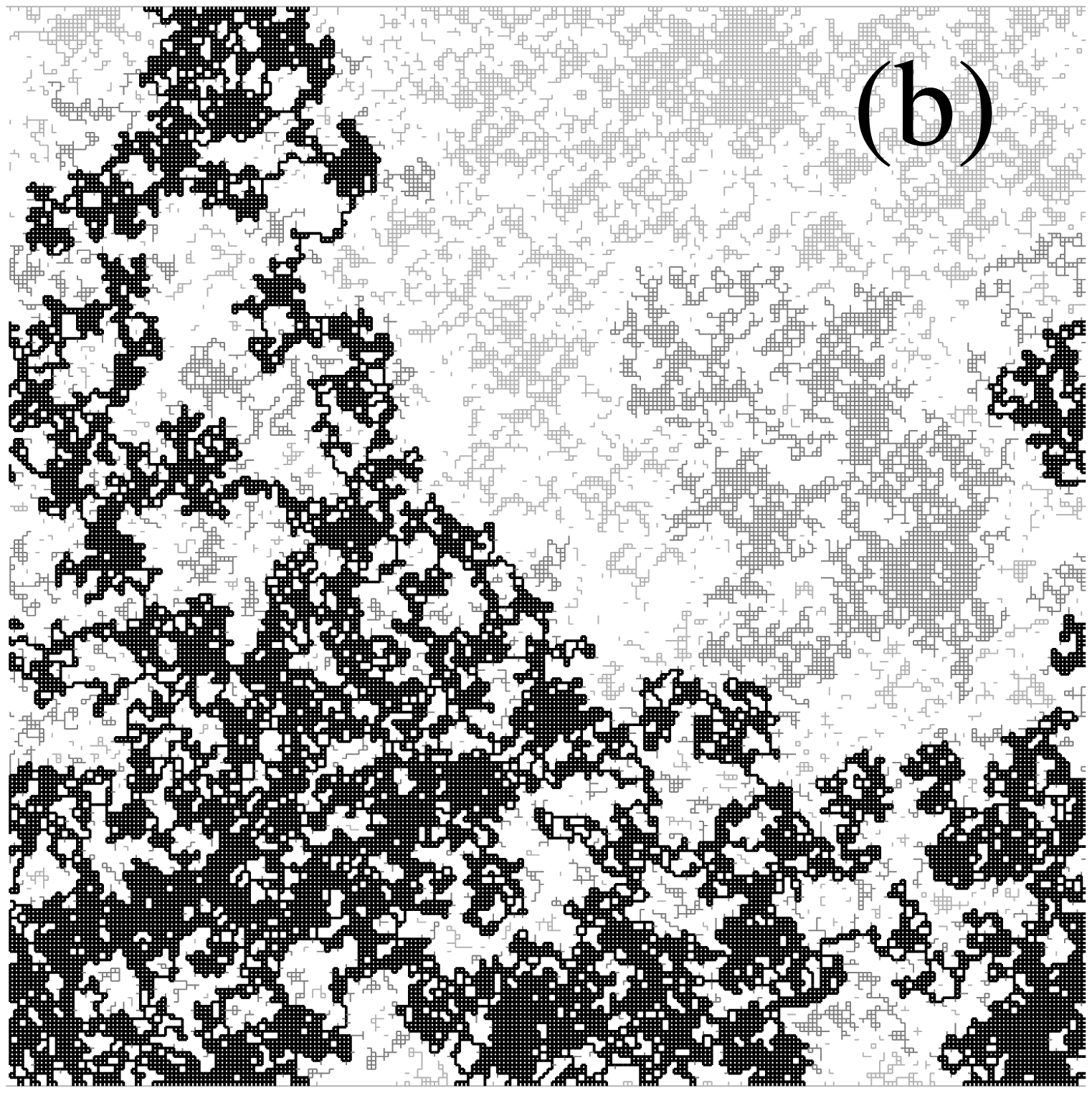}
\caption{(a) A typical percolation lattice $L=256$ for the uncorrelated
case. Heavy lines correspond to the backbone, gray lines to dangling ends,
and light gray lines to isolated ``islands'' (non-spanning clusters). 
(b) The same as in (a), but for the correlated case ($\gamma=0.5$).}
\end{figure}

\begin{figure}
\includegraphics[width=8.0cm]{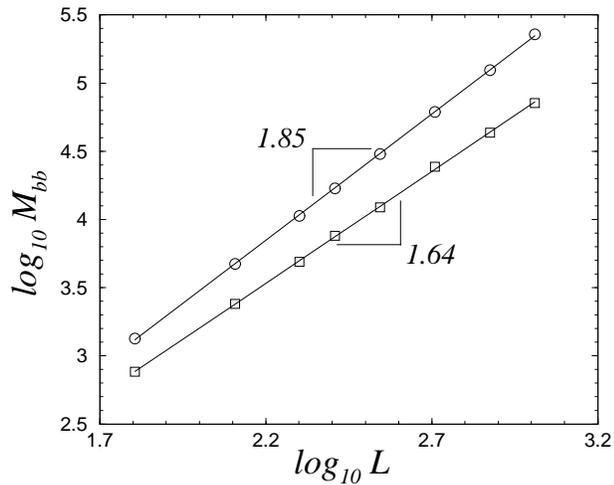}
\caption{Log-log plot of the backbone mass $M_{bb}$ versus the grid size $L$
for uncorrelated networks (squares) and correlated networks (circles). The
long-range correlated percolation structures have been generated with
$\gamma=0.5$. The solid lines are the least-square fits to the data with
slopes corresponding to the fractal dimensions of the respective backbones,
$d_{bb}$.}
\end{figure}

\begin{figure} 
\includegraphics[width=10.0cm]{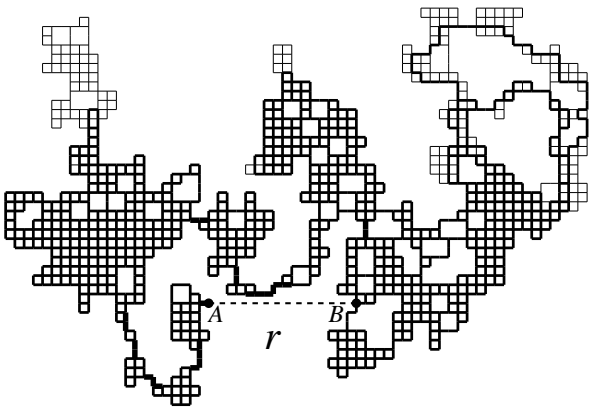}
\caption{The traveling paths of 10~000 tracers ($L=64$, $r=16$ and
$\gamma=0.5$). Heavy lines correspond to the bonds that receive more than
6000 tracers, medium lines to those that receive between 1 and 6000, and thin
lines to those that receive none.}
\end{figure}

\begin{figure}
\includegraphics[width=8.0cm]{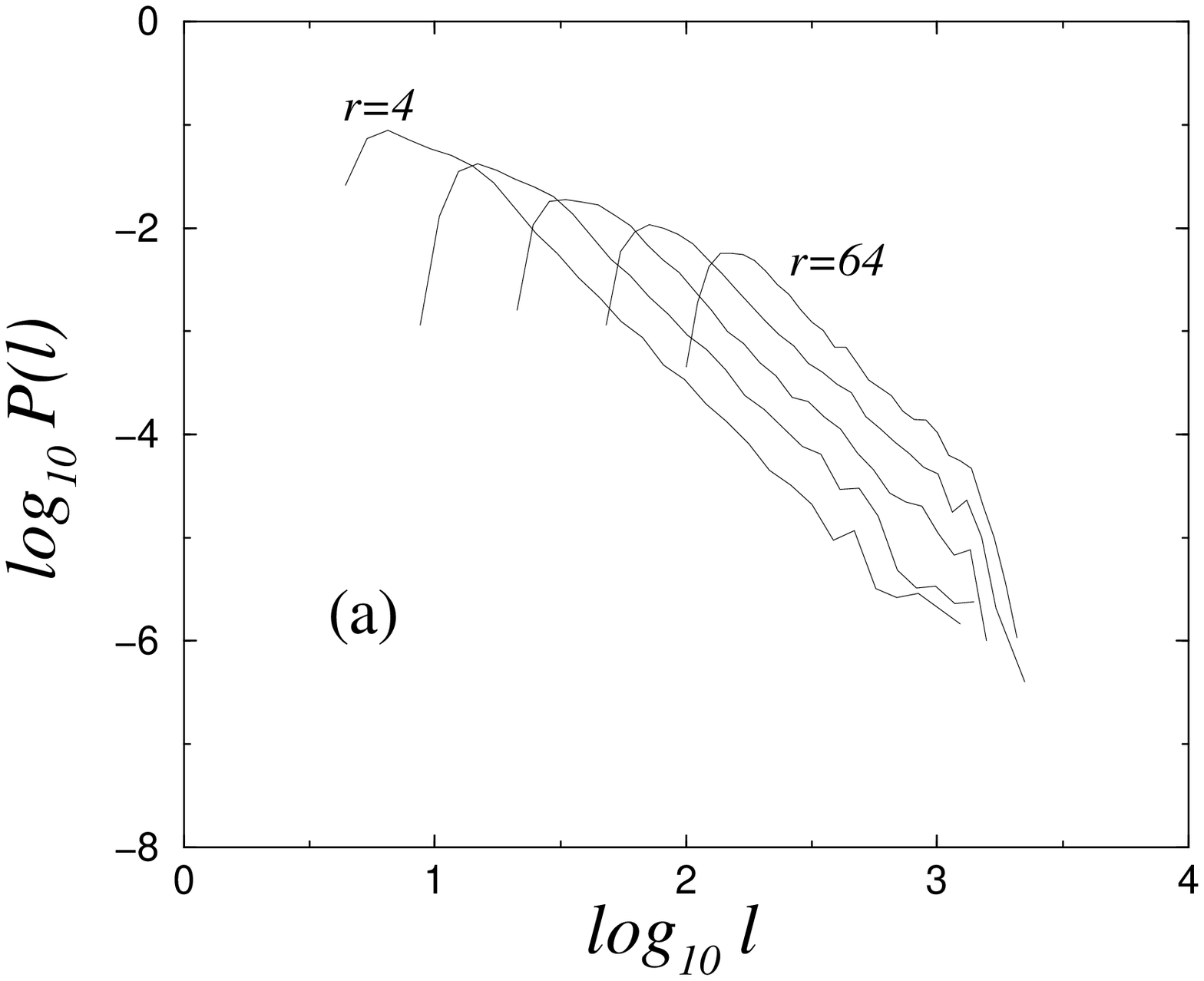}
\includegraphics[width=8.0cm]{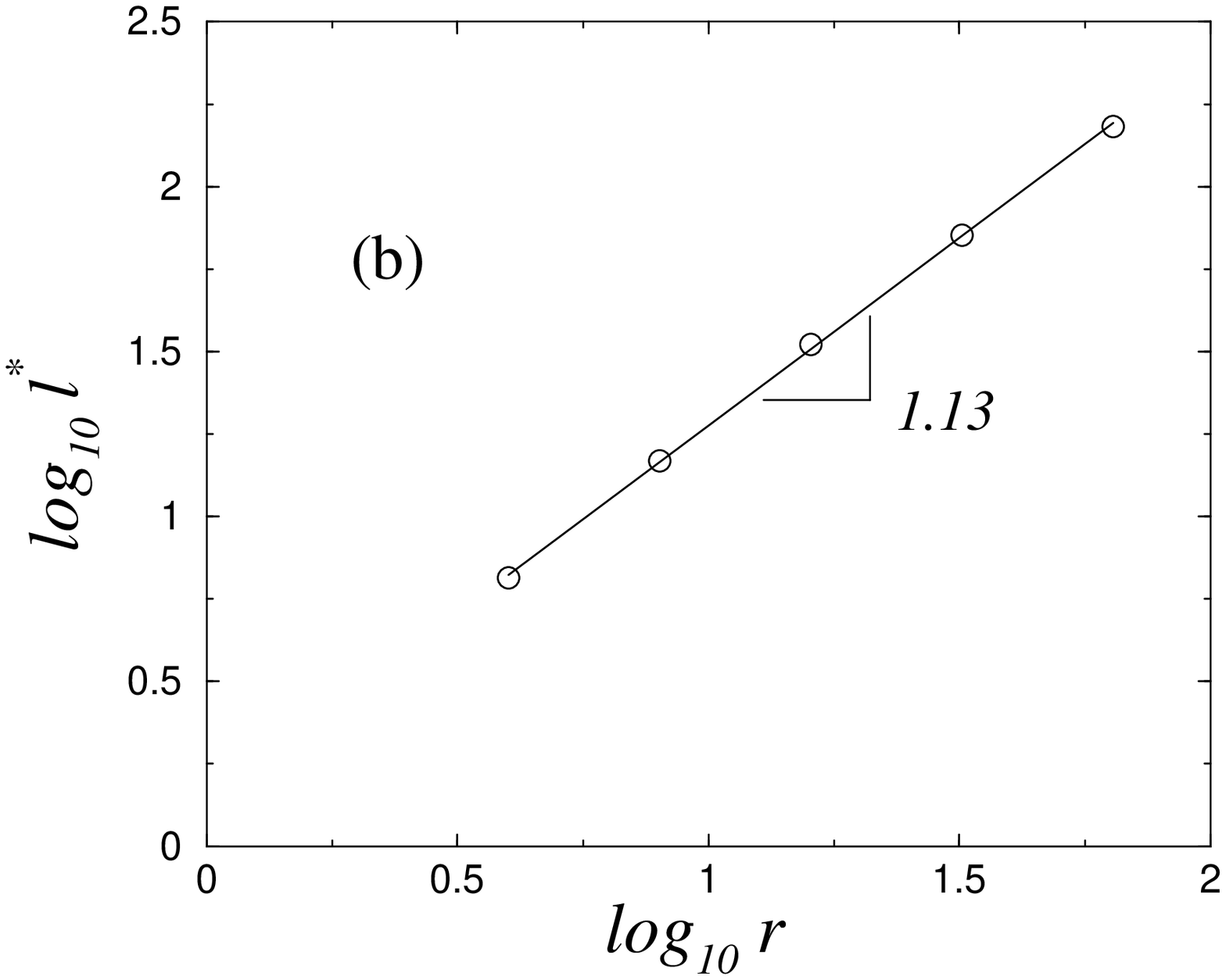}
\includegraphics[width=8.0cm]{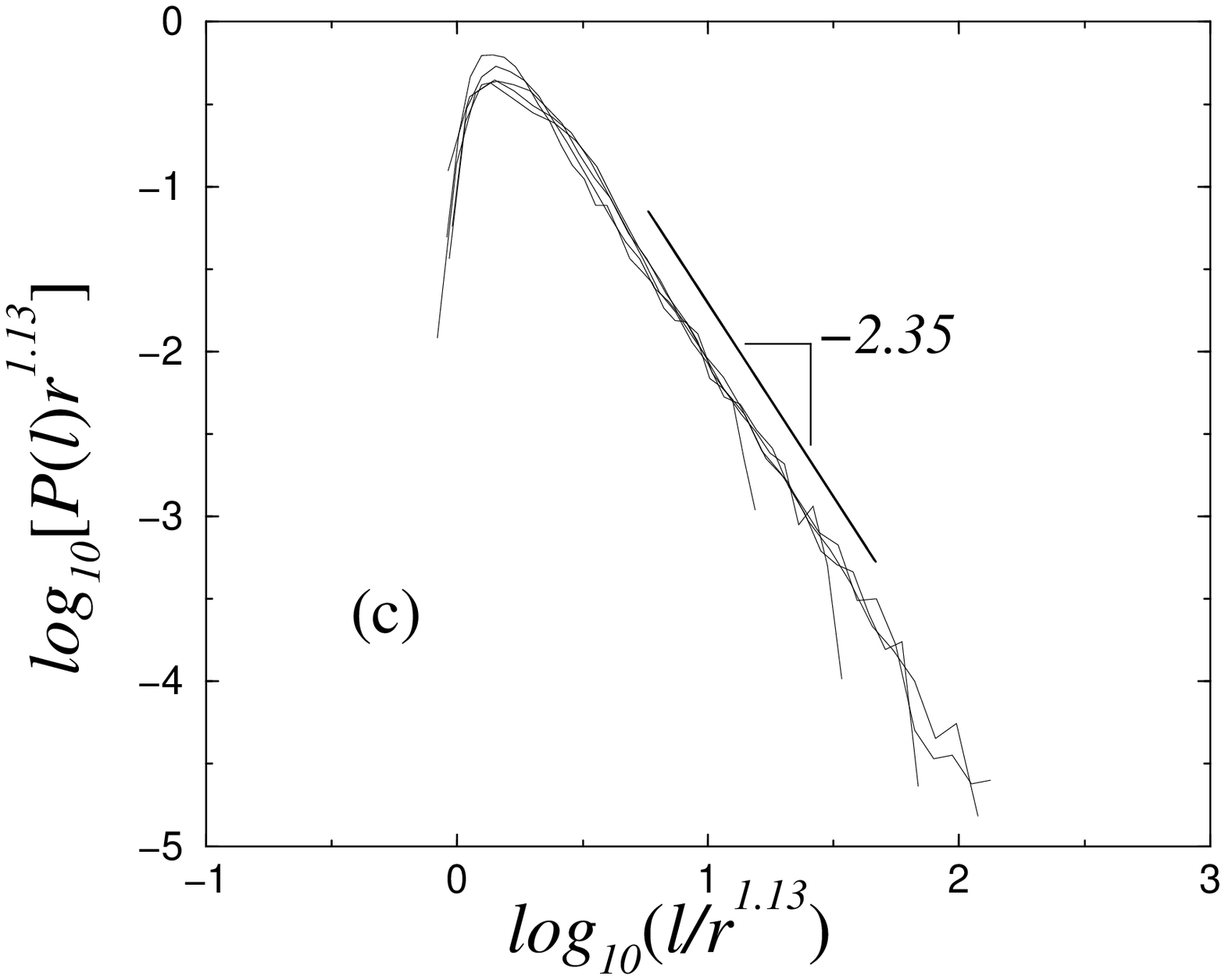}
\caption{(a) Log-log plot of traveling distance distribution $P(l)$ for
$\gamma=0.5$ and $r=4,8,16,32,64$. (b) Log-log plot of the most probable
value $l^\ast$ for traveling length versus the distance $r$. The
straight line is the least-squares fit to the data, $d_{l}=1.13 \pm
0.02$. (c) Data collapse obtained by rescaling $l$ with its
characteristic value $l^{*}\sim r^{1.13}$. The least-square fit to the
data in the scaling region gives $g_{l}= 2.35 \pm 0.05$.}
\end{figure}

\begin{figure}
\includegraphics[width=8.0cm]{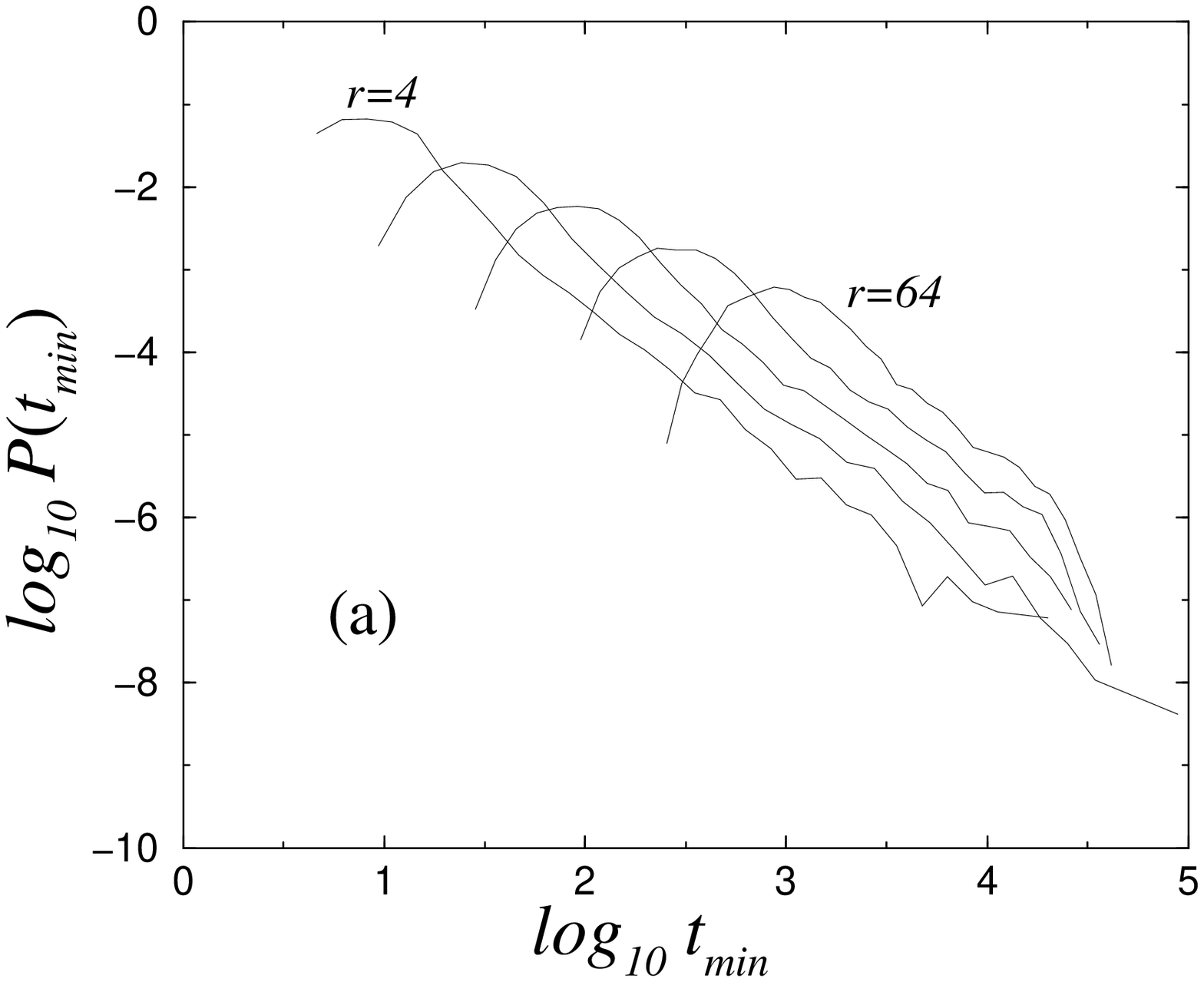}
\includegraphics[width=8.0cm]{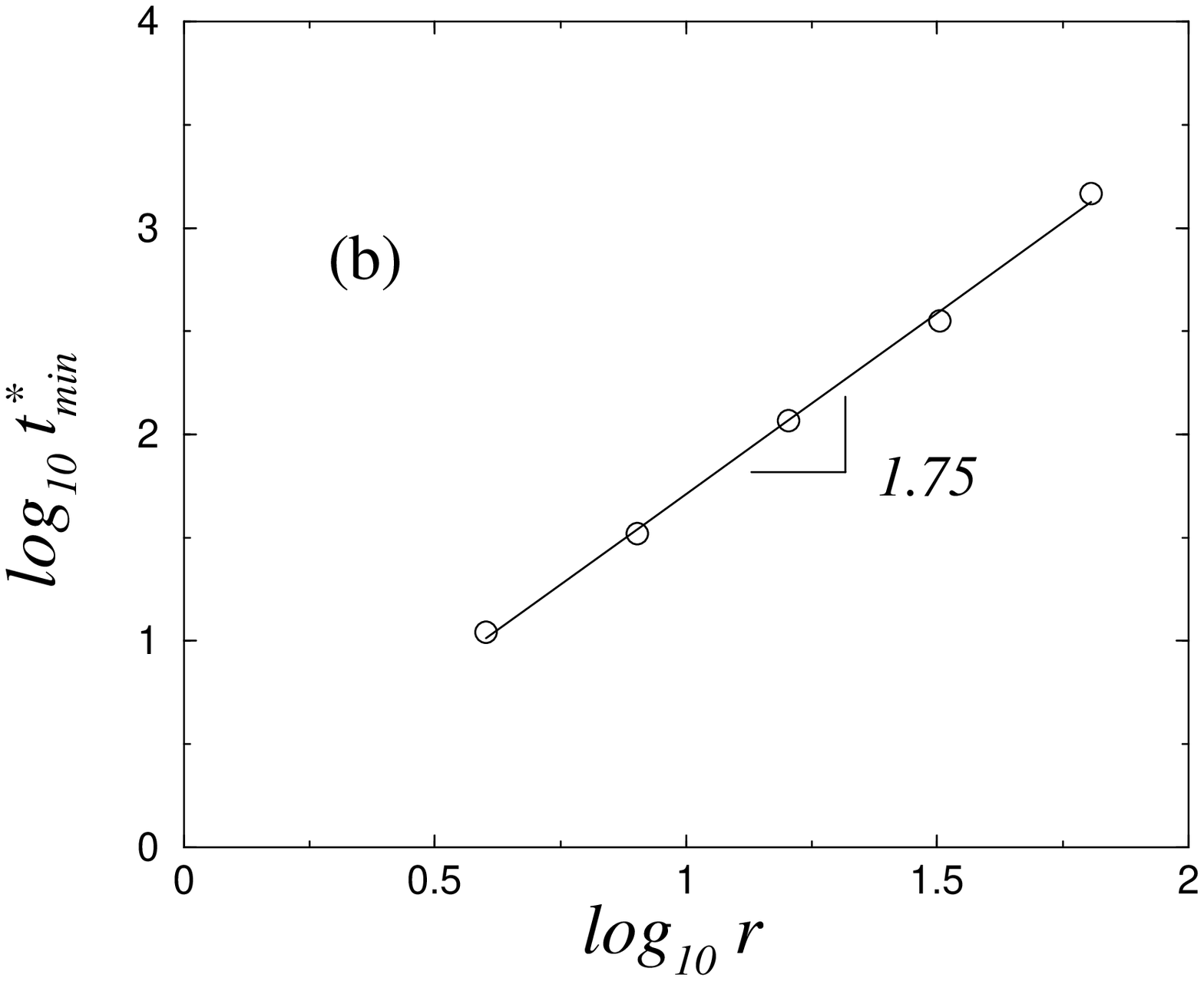}
\includegraphics[width=8.0cm]{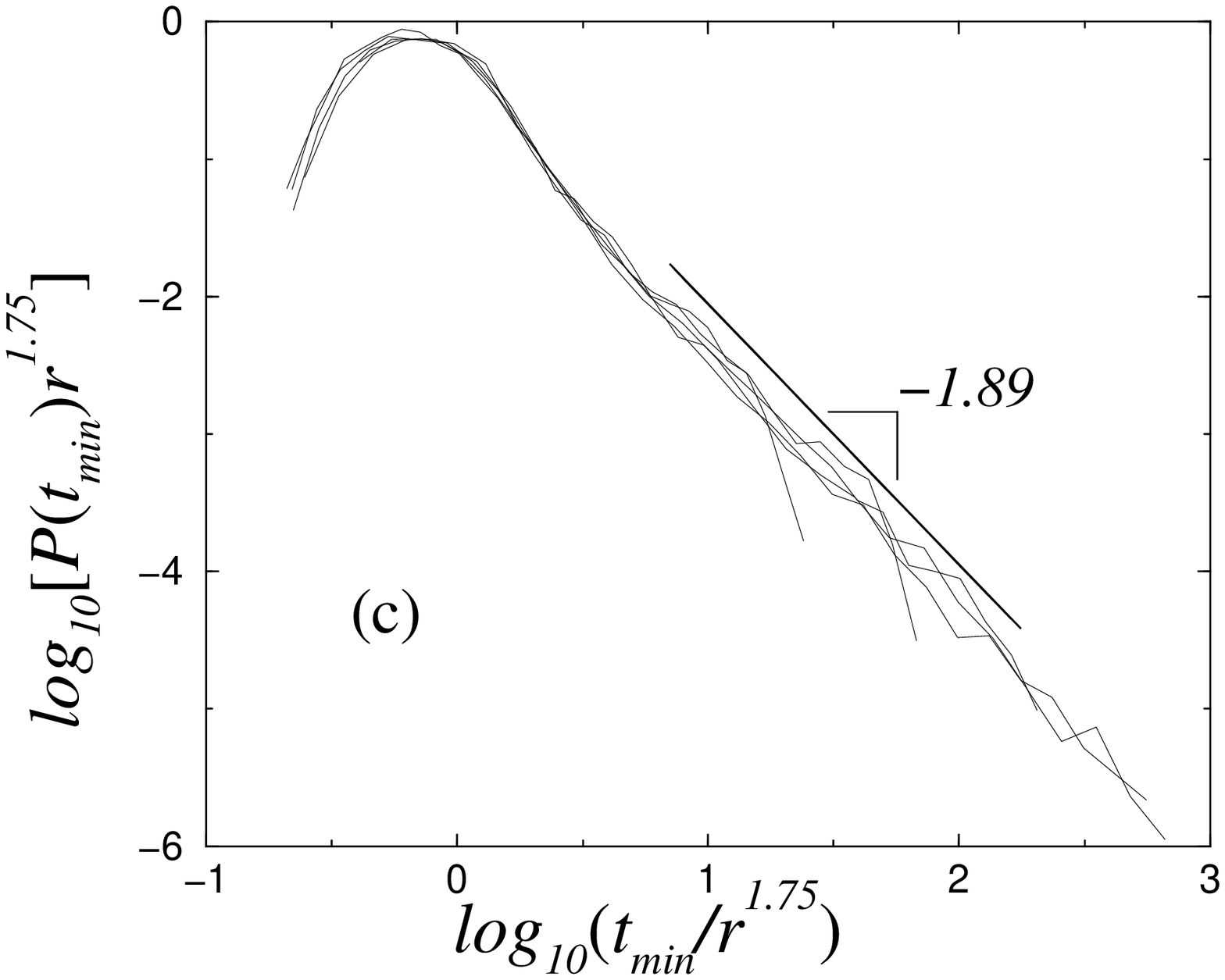}
\caption{(a) Log-log plot of the minimum traveling time distribution
$P(t_{min})$ for $\gamma =0.5$ and $r=4,8,16,32,64$. (b) Log-log plot of
the most probable values for the minimal traveling time $t_{min}$ versus
the distance $r$. The straight line is the least-square fit to the data,
with the number indicating the slope, $d_{t}=1.75 \pm 0.03$. (c) Data
collapse obtained by rescaling $t_{min}$ with its characteristic time
$t^{*}_{min}\sim r^{1.75}$. The least-square fit to the data in the scaling
region gives $g_{t}=1.89 \pm 0.04$.}
\end{figure}


\begin{thebibliography}{99}

\bibitem{Bear72} J. Bear,  \textit{Dynamics of Fluids in Porous
Materials} (Elsevier, New York, 1972).

\bibitem{Dullien79} F. A. Dullien, \textit{Porous Media - Fluid
Transport and Pore Structure} (Academic, New York, 1979).

\bibitem{Sahimi95} M. Sahimi, \textit{Flow and Transport in Porous
Media and Fractured Rock} (VCH, Boston, 1995).

\bibitem{Ben-Avraham00} D. Ben-Avraham and S. Havlin, \textit{Diffusion
and Reactions in Fractals and Disordered Systems} (Cambridge
University Press, Cambridge, 2000).

\bibitem{King90} P. R. King, in \textit{North Sea Oil and Gas
Reservoirs III}, ed. A. T. Buller \textit {et al}. (Graham and
Trotman, London, 1990).

\bibitem{Stauffer94} D. Stauffer and A. Aharony, \textit{Introduction to
Percolation Theory} (Taylor \& Francis, Philadelphia, 1994).

\bibitem{Bunde96} \textit{Fractals and Disordered Systems} 2nd ed.,
edited by A. Bunde and S. Havlin (Springer-Verlag, New York, 1996).

\bibitem{Ambegaokar71} V. Ambegaokar, B. I. Halperin, and J. S. Langer
Phys. Rev. B {\bf 4}, 2612 (1971).

\bibitem{Katz86} A. J. Katz and A. H. Thompson, Phys. Rev. B {\bf 34},
8179 (1986); J. Geophys. Res. B {\bf 92}, 599 (1987).

\bibitem{Murat86} M. Murat and A. Aharony, Phys. Rev. Lett. {\bf
57}, 1875 (1986).

\bibitem{Tian99} J. P. Tian and K. L. Yao, Phys. Lett. A {\bf 251},
259 (1999)

\bibitem{Lee99} Y. Lee, J. S. Andrade, Jr., S. V. Buldyrev,
N. V. Dokholyan, S. Havlin, P. R. King, G. Paul, and H. E. Stanley,
Phys. Rev. E. {\bf 60} , 3425 (1999).

\bibitem{Andrade00} J. S. Andrade, Jr., S. V. Buldyrev, N. V. Dokholyan,
S. Havlin, P. R. King, Y. Lee, G. Paul, and H. E. Stanley,
Phys. Rev. E. {\bf 62} , 8270 (2000).
 
\bibitem{Havlin87} S. Havlin and D. Ben-Avraham, Adv. Phys. {\bf
36}, 695 (1987).

\bibitem{Dokholyan98} N. V. Dokholyan, Y. Lee, S. V. Buldyrev,
S. Havlin, P. R. King, and H. E. Stanley, J. Stat. Phys. {\bf 93},
603 (1998).

\bibitem{Andrade01} J. S. Andrade, Jr., A. D. Ara\'ujo, S. V. Buldyrev,
S. Havlin, and H. E. Stanley, Phys. Rev. E. {\bf 63}, 051403 (2001).

\bibitem{Havlin88} S. Havlin, R. B. Selinger, M. Schwartz, H. E. Stanley,
and A. Bunde, Phys. Rev. Lett. {\bf 61}, 1438 (1988).

\bibitem{Havlin89} S. Havlin, M. Schwartz, R. B. Selinger, A. Bunde, and 
H. E. Stanley, Phys. Rev. A {\bf 40}, 1717 (1989).

\bibitem{Prakash92} S. Prakash, S. Havlin, M. Schwartz, and
H. E. Stanley, Phys. Rev. A. {\bf 46}, R1724 (1992).

\bibitem{Makse96a} H. A. Makse, S. Havlin, M. Schwartz, and
H. E. Stanley, Phys. Rev. E. {\bf 53}, 5445 (1996).

\bibitem{Makse96b} H. A. Makse, G. Davies, S. Havlin, P.-Ch. Ivanov,
P. R. King, and H. E. Stanley, Phys. Rev. E {\bf 54}, 3129 (1996).

\bibitem{Makse00} H. A. Makse, J. S. Andrade, Jr., and H. E. Stanley,
Phys. Rev. E. {\bf 61}, 583 (2000).

\bibitem{Gennes79} P. G. de Gennes, \textit{ Scaling Concepts in Polymer
Physics\/} (Cornell University Press, Ithaca, 1979).

\bibitem{King01} P. R. King, S. V. Buldyrev, N. V. Dokholyan, S. Havlin,
Y. Lee, G. Paul, H. E. Stanley, and N. Vandesteeg, Petrol. Geosci. {\bf
7}, S105 (2001). 

\bibitem{Kostek92} S. Kostek, L. M. Schwartz, and D. L. Johnson,
Phys. Rev. B {\bf 45}, 186 (1992).

\bibitem{Schwartz93} L. M. Schwartz, N. Martys, D. P. Bentz, E. J. Garboczi,
and S. Torquato, Phys. Rev. E {\bf 48}, 4584 (1993).

\bibitem{Martys94} N. Martys, S. Torquato, and D. P. Bentz, Phys. Rev. E
{\bf 50}, 403 (1994).

\bibitem{Koponen97} A. Koponen, M. Kataja, and J. Timonen, Phys. Rev. E 
{\bf 56}, 3319 (1997).

\bibitem{Andrade97} J. S. Andrade, Jr., M. P. Almeida, J. Mendes Filho,
S. Havlin, B. Suki, and H. E. Stanley, Phys. Rev. Lett. {\bf 79}, 3901 (1997).

\end{thebibliography}
\end{document}